%% file: main.tex
\documentclass[letterpaper, 10 pt, conference, keeplastbox]{ieeeconf}
\input{packages}
\title{\LARGE \bf 3-D Registration on Carotid Artery imaging data: MRI for different timesteps}
\input{affiliations}
\begin{document}
    \maketitle
    \thispagestyle{empty}
    \pagestyle{empty}
    \input{abstract}
    \input{introduction}
    \input{methods}
    \input{results}
    \input{conclusions}
    \begingroup
        \let\clearpage\relax
        \bibliographystyle{IEEEtran}
        %         \bibliography{EMBC2016}
        
    \endgroup
\end{document}

%% file: packages.tex
\IEEEoverridecommandlockouts
\usepackage{graphicx}
\usepackage{tikz}
\usepackage{subcaption}
\usepackage{booktabs}
\usepackage{flushend}
\usepackage{xr}
\usepackage{hyperref}
\usepackage{footnote}
\usetikzlibrary{shapes.geometric, arrows, fit}
\tikzstyle{box} = [rectangle, rounded corners, minimum width=2cm, minimum height=1cm,text centered, draw=black, fill=blue!30, align=center]
\tikzstyle{arrow} = [draw, -latex',thick]
\tikzstyle{c} = [rectangle, draw, dashed]
\graphicspath{{images/}}

%% file: affiliations.tex
\author{Paschalis~A.~Bizopoulos, \IEEEmembership{Student Member, IEEE}, Antonis~Sakellarios,  Lampros~K.~Michalis,\\ Dimitrios~D.~Koutsouris, \IEEEmembership{Senior Member, IEEE} and Dimitrios~I.~Fotiadis, \IEEEmembership{Senior Member, IEEE}
\thanks{Paschalis Bizopoulos and Dimitrios Koutsouris are with the Biomedical Engineering Laboratory, School of Electrical and Computer Engineering, National Technical University of Athens, GR 157 73 Zografou, Athens, Greece, paschalis.bizopoulos@gmail.com, dkoutsou@biomed.ntua.gr}%
\thanks{Antonis Sakellarios and Dimitrios I. Fotiadis are with the Unit of Medical Technology and Intelligent Information Systems, Dept. of Materials Science and Engineering, University of Ioannina, GR 451 10 Ioannina, Ioannina, Greece, ansakel@cc.uoi.gr, dimitris.fotiadis30@gmail.com}%
\thanks{Lampros Michalis is with the Michaelideion Cardiac Center, Department of Cardiology in Medical School, University of Ioannina, GR 451 10 Ioannina, Greece lamprosmihalis@gmail.com}%
\thanks{Dimitrios I. Fotiadis is with the Biomedical Research Division of the Institute of Molecular Biology and Biotechnology – FORTH, University Campus of Ioannina, GR 451 10, Ioannina, Greece.}%
}

%% file: abstract.tex
\begin{abstract}
A common problem which is faced by the researchers when dealing with arterial carotid imaging data is the registration of the geometrical structures between different imaging modalities or different timesteps. The use of the ``Patient Position'' DICOM field is not adequate to achieve accurate results due to the fact that the carotid artery is a relatively small structure and even imperceptible changes in patient position and/or direction make it difficult. While there is a wide range of simple/advanced registration techniques in the literature, there is a considerable number of studies which address the geometrical structure of the carotid artery without using any registration technique. On the other hand the existence of various registration techniques prohibits an objective comparison of the results using different registration techniques. In this paper we present a method for estimating the statistical significance that the choice of the registration technique has on the carotid geometry. One-Way Analysis of Variance~(ANOVA) showed that the p-values were \textless 0.0001 for the distances of the lumen from the centerline for both right and left carotids of the patient case that was studied.
\end{abstract}

%% file: introduction.tex
\section{INTRODUCTION}
Cardiovascular disease~(CVD) is one of the leading causes of disability and death worldwide~\cite{DAgostino2013}. Plaque vulnerability causes plaque rupture which is directly related with CVD. The evaluation and diagnosis of the current condition and the evolution of the plaque requires the assessment of the arterial geometry, thus it is crucial to monitor plaque changes in morphology during time.
\par
It is evident from the literature that accurate assessment of the changes in carotid geometry is vital for estimating the risk of plaque rupture. Holden~\cite{Holden2008} has generically reviewed the geometric transformations for non-rigid body registration. Phan et al.~\cite{Phan2012} have indepentently associated various Internal Carotid Artery~(ICA) geometry measures with ICA stenosis. A geometry related risk factor that has been thoroughly studied is Intima-Media Thickness (IMT). O'Leary et al.~\cite{OLeary1999} have concluded that increased IMT of the carotid artery is associated with increased risk of stroke in certain populations. Lorenz et al.~\cite{Lorenz2007} after conducting a meta-analysis found that Carotid IMT is a strong predictor of future vascular events. Nanayakkara et al.~\cite{Nanayakkara2008} have used a ``twist and bend'' model on ultrasound (US) images of carotid plaque obtained at different time points. Gupta et al.~\cite{Gupta2013} have proposed a hybrid technique for US registration using rigid transformation along with mutual information and Powell optimizer. Registration techniques have also been used in Dynamic Contrast-Enhanced MRI by Ramachandran et al.~\cite{Ramachandran2014} using a template-based squared-difference method.
\par
There exists also an increasing number of studies that tackle the problem of registration in carotid arteries between different modalities. Nanayakkara et al.~\cite{Nanayakkara2009} have also used the ``twist and bend'' model to combine Magnetic Resonance Imaging~(MRI) and ultrasound~(US) carotid images. Carvalho et al.~\cite{Carvalho2014} have introduced a semiautomatic method for registering free-hand B-Mode US and MRI of the carotid artery. Chiu et al.~\cite{Chiu2012} have developed a surface-based algorithm for US and MRI using iterative closest point~(ICP). Loeckx et al.~\cite{Loeckx2007} have proposed the use of non-rigid registration on Computed Tomography Angiography~(CTA) images.
\par
The objective of our study is to test the statistical significance of the use of two rigid registration techniques. The variable that was tested was the difference of the distance of the lumen wall from the centerline in the baseline and follow-up, with and without using a rigid registration technique. The rigid registration techniques tested are: one using the whole carotid artery and another using separately each one of the branches (common, internal and external).

%% file: methods.tex
\section{METHODS}
The image acquisition of the data that were used~\cite{Sakellarios2012} was Time-Of-Flight~(TOF) MRI, which is a technique that allows visualization of the flow within the carotid artery. Two timesteps of the left and right carotid from one patient were used, at baseline and at follow-up (6 months). 
\par
The first step of preprocessing consists of manual choosing a rectangular bounding box around the carotid artery. Then the wall of the lumen of the carotid artery was reconstructed using a manually chosen intensity threshold, without loss of generality. All the visualizations were made using the functionalities of our previously published software~\cite{Bizopoulos2014}.
\par
Three cases were compared. In the first one, the follow-up remained unchanged. In the second one intensity-based monomodal registration was applied on the volume as a whole~\cite{MATLAB2015} using rigid transformation. The baseline volume was used as the fixed reference and the follow-up as the moving target. In the third case the common, internal and external carotid branches were split using a cut-off plane to a manually chosen height. 
\par
\begin{figure}[thpb]
    \centering
    \includegraphics[scale=1.1]{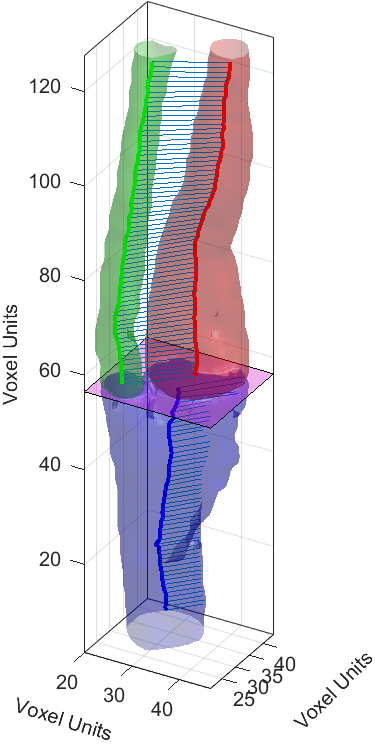}
    \caption{The blue, red and green isosurfaces depict the common, external and internal carotid branches respectively. The same color rule applies for the thick lines that depict the centerlines of the branches. The thin blue lines that connect the points of the centerlines of the internal and external and the points of the centerline of the common with the common wall, depict the base for the angles and the magenta horizontal planes separate the common branch with the external and internal branches.}
    \label{fig:BaselineSeparate}
\end{figure}

In all cases the vertical position of the registered follow-up was manually adjusted in such a way that the apex of the registered follow-up coincided with the apex of the fixed reference to avoid divergence of the registration's optimization. Then, 100 horizontal cut-off planes were used (46 for common, 54 for internal and external). The intersected curves were interpolated in 100 points and in-between aligned. The cases are depicted in Fig.\ref{fig:flowchartOfTheProcedure} which provides the methodology of our approach. The set of linear segments that connect the points of the centerline of the external branch and the points of the centerline of the internal branch were used as reference for the angles, as it is shown in Fig.\ref{fig:BaselineSeparate}. In Fig.\ref{fig:registration} the three overlapping isosurfaces are depicted with corresponding colors for each branch and cross-section of the horizontal planes.

\input{flowchart}
\input{registrationFigure}

%% file: flowchart.tex
\begin{figure}[thpb]
    \centering
    \begin{tikzpicture}[node distance=1cm]
        \node (boundingBox) [box] {Bounding Box};
        \node (intensitySegmentation) [box, below of=boundingBox, node distance=1.7cm] {Intensity Segmentation};
        
        \node (noChange) [box, below left of=intensitySegmentation, node distance=3.3cm] {No Change};

        \node (rigidRegistrationUnified) [box, below of=intensitySegmentation, node distance=3cm] {Rigid\\ Monomodal\\ Registration};

        \node (branchSplitting) [box, below right of=intensitySegmentation, node distance=3.3cm] {Carotid\\ Branches\\ Split};
        
        \node (rigidRegistrationSeparated) [box, below of=branchSplitting, node distance=2cm] {Rigid\\ Monomodal\\ Registration};
        
        \node (apexCoincide) [box, below of=rigidRegistrationUnified, node distance=2cm] {Manual\\ Apex\\ Coincide};
        
        \node (planeCut) [box, below of=apexCoincide, node distance=1.9cm] {Horizontal\\ Cutting\\ Planes};
        
        \node (compareResults) [box, below of=planeCut, node distance=1.7cm] {Compare Results};
        
        \node[c,fit=(boundingBox)(intensitySegmentation), label={above:Preprocessing}] (container) {};
        
        \node[c,fit=(noChange), label={above:No Registration}] (container) {};
                
        \node[c,fit=(rigidRegistrationUnified), label={above:No Branch Split}] (container) {};
        
        \node[c,fit=(branchSplitting)(rigidRegistrationSeparated), label={above:With Branch Split}] (container) {};
        
        \path [arrow] (boundingBox) -- (intensitySegmentation);

        \path [arrow] (intensitySegmentation) -- (noChange);
        \path [arrow] (noChange) -- (apexCoincide);

        \path [arrow] (intensitySegmentation) -- (rigidRegistrationUnified);
        \path [arrow] (rigidRegistrationUnified) -- (apexCoincide);

        \path [arrow] (intensitySegmentation) -- (branchSplitting);
        \path [arrow] (branchSplitting) -- (rigidRegistrationSeparated);
        \path [arrow] (rigidRegistrationSeparated) -- (apexCoincide);
        
        \path [arrow] (apexCoincide) -- (planeCut);
        \path [arrow] (planeCut) -- (compareResults);

    \end{tikzpicture}
    \caption{Flowchart of our approach. The blue boxes depict a process, the arrows depict the flow and the dashed rectangles depict grouped processes. The grouped processes names are located on the top of each dashed rectangle.}
    \label{fig:flowchartOfTheProcedure}
\end{figure}

%% file: registrationFigure.tex
\begin{figure*}
    \centering
    \begin{subfigure}{.3\textwidth}
        \centering
        \includegraphics[scale=0.77]{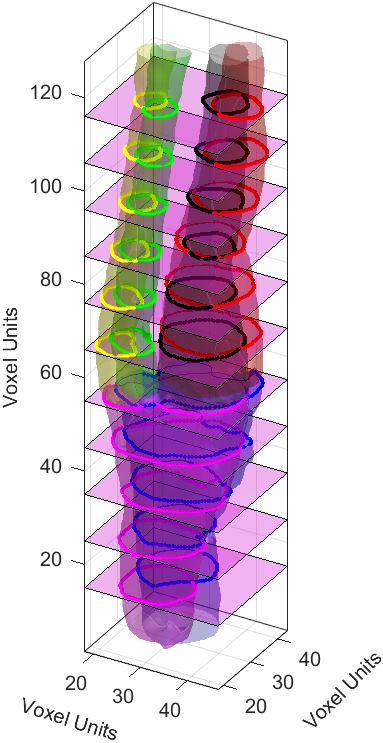}
        \caption{Without Registration.}
        \label{fig:BaselineWithFollowUpSeparate}
    \end{subfigure}
    \qquad
    \begin{subfigure}{.3\textwidth}
        \centering
        \includegraphics[scale=0.77]{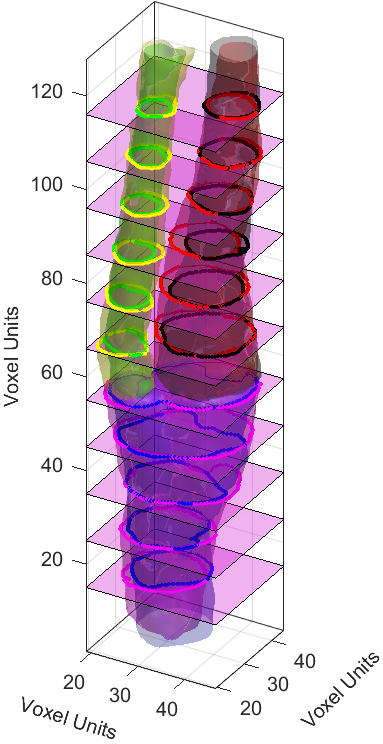}
        \caption{With Registration.}
        \label{fig:BaselineWithUnifiedRegisteredFollowUpSeparate}
    \end{subfigure}
    \qquad
    \begin{subfigure}{.3\textwidth}
        \centering
        \includegraphics[scale=0.77]{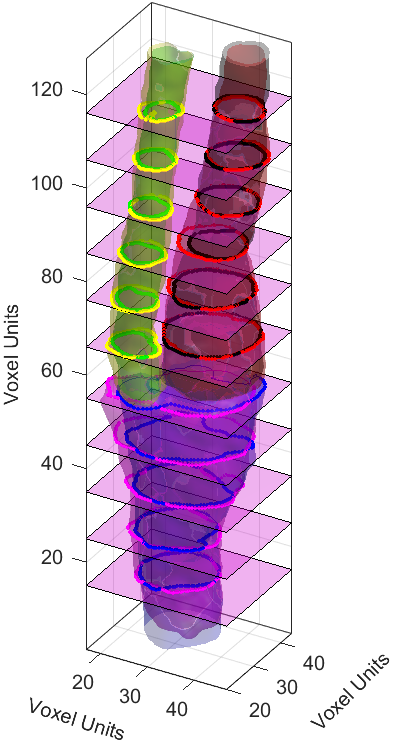}
        \caption{With Registration on each branch.}
        \label{fig:BaselineWithRegisteredFollowUpSeparate}
    \end{subfigure}
    \caption{The blue, red and green isosurfaces depict the common, external and internal carotid branches, respectively, at the baseline. The magenta, black and yellow isosurfaces depict the common, external and internal carotid branches, respectively, at the follow-up without registration in (a), with registration in (b) and with registration for each branch separately in (c). The closed curves with the previously described colors depict cross-sections of the carotid artery with the horizontal planes. Not all planes are shown for visualization purposes.}
    \label{fig:registration}
\end{figure*}

%% file: results.tex
\section{RESULTS}
\begin{figure}[thpb]
    \centering
    \includegraphics[scale=0.28]{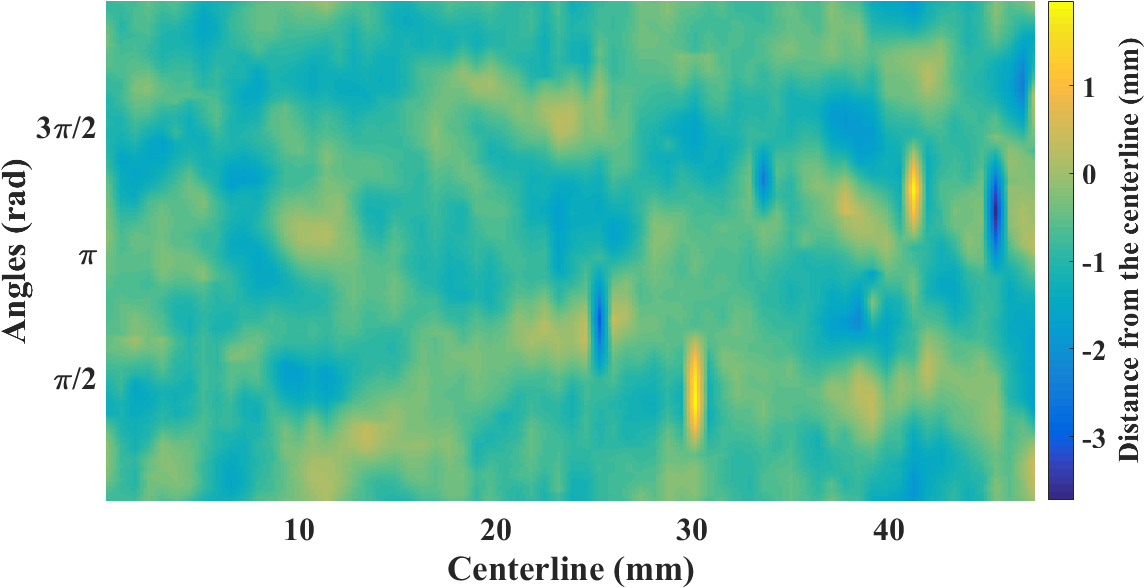}
    \caption{Spread-out surf plot of the difference of the distance of wall of the unchanged follow-up from the centerline and the unified registered external branch of the right carotid.}
    \label{fig:surf}
\end{figure}

The distances from the centerline for each case and each branch were subtracted from the distances from the centerline at the baseline separately. An indicative spread-out plot is shown in Fig.\ref{fig:surf}. The plot has been interpolated using 4 times the original points in order to smoothen the surface.

The mean absolute difference of distances of the tested variable were calculated for the two timesteps for both registration techniques and the results are shown in Table.~\ref{meanAbs}. The results were grouped into three groups one for each branch and then each one of the groups was analyzed using \mbox{One-Way Analysis of Variance}~(ANOVA). The p-values for all the three groups were \textless 0.0001 with 4600 and 3700 samples for the common branches and 5400 and 7300 samples for the internal and external branches, for the right and left carotid, respectively. Therefore, the choice of a registration method (or the absence of it) on the specific carotid arteries that were studied statistically influenced the results. By visually comparing the results of the two registrations techniques it is observed that the one using the separated branches failed to fit the common artery near the bifurcation region related with the one using the artery as a whole.

\begin{table}[thpb]
    \centering
    \caption{Mean Absolute distance difference for right and left carotid (in voxel units).}
    \label{meanAbs}
    \begin{tabular}{@{}ccccccc@{}}
        \toprule
        & \multicolumn{3}{c}{Unified}  & \multicolumn{3}{c}{Separate Branches} \\ \midrule
        & Common & Internal & External & Common    & Internal    & External    \\ \midrule
        R & 1.18 & 0.65 & 0.72 & 1.19 & 0.96 & 0.77\\
        L & 4.54 & 0.38 & 0.54 & 3.69 & 0.75 & 1.07       
    \end{tabular}
\end{table}

%% file: conclusions.tex
\section{CONCLUSIONS}
In this study it was demonstrated that the choice of the \mbox{3-D} registration technique on carotid arteries is important and it has considerable effect on the geometrically related results. One of the future work is to test our approach using a considerable amount of data in order to increase the statistical significance of the results. Moreover, the results need to be validated using a golden standard either by using artificially generated data (on which the golden standard is known with maximum accuracy) or either by creating a method that allows the expert to create annotations on these type of data. The described approach can also be applied on multi-modal data.